\documentclass[allclo,superscriptaddress,eqsecnum,amsfonts,showpacs]{revtex4}
\usepackage{epsfig}

\newcommand{\be}{\begin{equation}}
\newcommand{\ee}{\end{equation}}
\newcommand{\bea}{\begin{eqnarray}}
\newcommand{\eea}{\end{eqnarray}}

\newcommand{\ep}{i\varepsilon}
\newcommand{\nn}{\nonumber}

\begin{document}

\preprint{ \parbox{1.5in}{\leftline{hep-th/??????}}}

\title{Behaviour of propagator and quark confinement}

\author{Vladimir ~\v{S}auli}
\affiliation{CFTP and Dept. of Phys.,
IST, Av. Rovisco Pais, 1049-001 Lisbon,
Portugal }
\affiliation{Dept. of Theor. Phys., INP, \v{R}e\v{z} near Prague, AV\v{C}R}

\begin{abstract}

The  propagator of confined quarks is calculated for timelike momenta  by transforming Minkowski Greens functions to the Temporal Euclidean space.
Based on the framework of the Schwinger-Dyson equations the QCD quark propagator is obtained  in  two  approximations which  differ by assuming behaviour  of  gluon propagator. In both studied cases we get  universal result for the light quarks: The quark mass function becomes complex bellow expected perturbative threshold, the obtained absolute value of the infrared mass is  $M\simeq \Lambda_{QCD} $ with the infrared  phase 
$\simeq {\pi\over 2}$. Permanent confinement of quarks is maintained by generation of the complex mass function which  prevents a real pole in the propagator. We will show that 
timelike dynamical Chiral Symmetry Breaking (CSB) solution  is  approximately, but non-trivially determined by the solution of gap equation in the standard Euclidean space.

\end{abstract}

%\pacs{11.15.Ha,87.10.Rt,12.20-m,25.75.Nq}
\maketitle
%

%%%%%%%%%%%%%%%%%%%%%%%%%%%%%%%%%%%%%%%%%%%%%%%%%%%%%%%%%%%%%%%%%%%%%%%%%%%%%%%%%%%%%
\section{Introduction}

Quark confinement in Quantum Chromodynamics (QCD) is theoretically unsolved phenomenon of longstanding interest \cite{MANDL1975,MANDL1976,POLYAK1977,THOFT1978,JCM1980,SUYO1989,AG2007,NEMCOURI}. QCD as the strong coupling theory is not easily tractable,
thus the confinement of colored object has been questioned in various indirect ways. Considering static infinitely heavy color sources the linearly rising potential was predicted and observed by simulation on the Euclidean lattice \cite{BALI2000}.

The alternative to the lattice theory is the continuous framework of
 the Schwinger-Dyson equations (SDEs) which in principle could provide a unique powerful tool for the nonperturbative QCD study (for a reviews see\cite{ROBERTS,ROBMAR2003,AS2001,BIPA2008}).
While it is naturally expected that  the SDEs Euclidean formulation  should approximately corresponds to the continuum limit of  lattice data whenever they are available, however the real advantage of SDEs approach is the possibility to explore directly all Minkowski space. To that point, based on the spectral representation, the problem has been formulated in Minkowski space and the results have been actually obtained  for a weak or  a medium coupling quantum field theory (for a review see \cite{SAULI2006}). Actually for  several cases of toy models the true equivalence between Minkowski  Greens functions evaluated at spacelike  momenta and the calculated in the Euclidean space form the beginning has been explicitly checked by the numerical solutions. For gluodynamics (QCD without quarks) the problem was  formulated  by using covariant Gauge/Pinch technique method time ago \cite{CORNWALL}, however never solved in Minkowski space in practise. Recently, we have no idea how much the analytical assumptions are justified for gluons, however we have no doubts that the usual analyticity  is too strong assumption for the quark propagator \cite{SABIAD}. To get  chiral symmetry breaking with an analytical quark propagator, i.e. which has a standard branch cut on the positive $p^2$ semi-axis, is very likely impossible - the quark propagator doesn't satisfy Lehmann representation.

 Direct solution of SDEs for strong coupling theory system in Minkowski space is still very complicated task and as we discuss above it requires an additional assumptions. To get rid of part of the mentioned weaknesses (e.g. assumption of the spectral representation) the SDEs has been recently formulated in Temporal Euclidean (ET) space \cite{SABA2008,SABA2009}, providing us with the solution for Greens function for the timelike, but Euclidean momenta. In this case we have the Euclidean metric $ p_{ET}^2=\sum_{i=1}^4 p_i^2$ 
\be \label{pres}
p=(p_1,p_2,p_3,p_4)^{[ET]}=  (p_0,-ip_1,-ip_2,-ip_3)^{[M]} \, ,
\ee
where Eq. (\ref{pres} ) explicitly shows the prescription between the components of Temporal Euclidean -ET- and Minkowski -M- fourvector. Recall here,  the prescription for the  measure $d^4p_M=id^4_{ET}$ with real $\pm\infty$ boundaries in a momentum loop integrals. Likewise for the standard -time component- Wick rotation \cite{WICK} the equivalence between ET space and Minkowski subspace is the assumption. Here, it is noteworthy that it was recently proved in \cite{SABA2009} that the
ladder approximation of Minkowski QED2+1  is exactly equivalent to the formulation of the problem in $ET$ space.

If the full fermion propagator has no mass singularity in the timelike region, it can never be on-shell and thus never observed as a free particle \cite{CORN1980,GRIBOV91,ROWIKR1992,MARIS1995}.
In our solution the imaginary part of the mass function is automatically generated for a coupling strong enough.
In the paper \cite{SABA2008} such result was firstly  obtained for explicitly massive quark
with bare current mass $m\sim \Lambda_{QCD}$. At  that time the authors of \cite{SABA2008} were not able to obtain the results for an arbitrary quark mass because of numerical obstacles accompanying their specific model. Therefore the main purpose of our paper is to present a models and techniques that exhibit good numerical stability for  all quarks flavors, e.g.  for the light $u,d$ quarks and hypothetical massless case as well. We will show that the light quark mass is the complex functions with relatively tiny  real parts in the infrared. The infrared complex phase is approximately given by the ratio of the current and the infrared (constituent) quark mass.
Quite interestingly, described complexification  is  universal and it doesn't depend on the details of the interaction kernel. As we will show explicitly the qualitative feature of observed complexification is independent  on  the (un)presence of (pole) type singularity  in the kernel.

As an another interesting model possibility we decreased the infrared running coupling and we study confinement and CBS phenomena near the phase transition, where generated mass is particularly small when compare to the scale of confinement, $M<<\Lambda$. The observed results are relevant for the Technicolor models, eg. to walking Technicolors wherein  large number of fermions makes the strong coupling softer in the infrared (originally these models have been developed to avoid flavor changing  neutral current, however they have theirs own interest).

 After the introduction of the models details in the next Section the numerical results are presented in the Section III. In this section we discuss the  relation  with the standard Euclidean solution. In the Section IV we summarize and conclude.

\section{The models- dressed ladder approximation of quark SDE}

The quark gap equation we solved is the  rainbow ladder approximation of the full quark SDE. 
In this Section we describe the details of the models employed here.
In the ladder approximation the quark SDE reads
\bea \label{gap}
S^{-1}(p)&=&S^{-1}_0(p)-\Sigma(p) \, ,
\nn \\
\Sigma(p)&=&i C_A g^2\int\frac{d^4q}{(2\pi)^4}\gamma_{\alpha}
G^{\alpha\beta}(p-q)S(q)\gamma_{\beta} \, ,
\eea
where the full gluon-quark-antiquark vertex of the exact SDE has been replaced by the standard four dimensional Dirac matrix, which obey anticomutation relation    $\left\{\gamma_{\mu},\gamma_{\nu}\right\}=2g_{\mu\nu}$ (we use the Minkowski metric $g_{\mu\nu}=diag(1,-1,-1,-1)$) and where  $C_A=T_a T_a=4/3$ for $SU(3)$ group and $G^{\alpha\beta}$ is the gluon propagator. The quark propagator $S$ can be parametrized
by two scalar functions conventionally like
\be
S(p)=S_v(p)\not p+S_s(p)=\frac{1}{\not p A(p)-B(p)}\, . 
\ee
The renormgroup invariant mass function is  defined as $M=B/A$, the physical mass $M_p$ could
be identified by the pole position of $S$ ,i.e. by the solution $ M^2(M_p^2)=M_p^2$ in unconfining theory. 

We assume these functions are complex, therefore it is convenient to parametrized them in  the following way
\bea
S_s(x)&=&\frac{B(k)}{A^2(k)k^2-B^2(k)}
\nn \\
&=&\frac{R_B\left[(R_A^2-\Gamma_A^2)k^2-R_B^2-\Gamma_B^2\right]+2R_A\Gamma_B\Gamma_A\,k^2}{D}
\nn \\
&+&i\, \frac{\Gamma_B\left[(R_A^2-\Gamma_A^2)k^2+R_B^2+\Gamma_B^2\right]-2R_BR_A\Gamma_A\,k^2}{D} \, ,
\label{ss}
 \\
S_v(k)=&=&\frac{A(k)}{A^2(k)k^2-B^2(k)}
\nn \\
&=&\frac{R_A\left[(R_A^2+\Gamma_A^2)k^2-R_B^2+\Gamma_B^2\right]-2R_B\Gamma_A\Gamma_B}{D}
\nn \\
&+&i\, \frac{\Gamma_A\left[-(R_A^2+\Gamma_A^2)k^2
-R_B^2+\Gamma_B^2\right]+2R_A R_B\Gamma_B}{D}\, \, ,
\label{sv}
\eea
where $R_A,R_B$ $(\Gamma_A,\Gamma_B)$ are the real (imaginary) parts of the functions $A,B$
and the denominator $D$ reads
\be
D=([R_A^2-\Gamma_A^2] k^2-[R_B^2-\Gamma_B^2])^2+4(\Gamma_A R_A-\Gamma_B B)^2 \,.
\ee

The last missing ingredient which completes our considered gap equation is the gluon propagator. With the exception of the large timelike momenta $p^2>\Lambda^2_{QCD}$, where the effective coupling is small and the result is  available by the analytical continuation of the Euclidean perturbation QCD,  gluon propagator is basically unknown function for timelike momenta. Most of the last decades  nonperturbative  studies were devoted  to the quark SDE in Landau gauge, thus for  a possible comparison we will work in this gauge as well. The Landau gauge Gluon propagator
is completely transverse
\be
G_{\mu\nu}=\frac{-g_{\mu\nu}+(1-\xi)\frac{k_{\mu}k_{\nu}}{k^2}}{k^2}G(k^2) \, ,
\ee
and is fully determined  by the gluon form factor $G$ which we will model as it is described bellow.

\subsection{Model I.}

 The model I. is based on a simple generalization of the perturbative one loop result for 
the gluon form factor. For this purpose we consider the kernel 
(in fact the effective product of  gluon propagator and $\gamma_ {\mu}$ part of the vertex function) such that it has a standard single pole and the the function $G$ is taken as 
\bea
\frac{g^2}{4\pi}G(q^2)=\frac{4\pi/\beta}{\frac{1}{2}ln\left[e+\left(\frac{q^2}{\Lambda^2} \,\right)^2\right]} \label{piva}
\eea
 where $\beta$ in (\ref{repa}) represents the beta function coefficient, for which we take $4\pi/\beta=1$ (recall,  $4\pi/\beta=1.396$ for three active quarks in perturbative QCD).
The prefactor is adjusted  in a way that Eq. (\ref{piva})  behaves as QCD running  coupling at ultraviolet, i.e.
\be \label{prop}
\frac{g^2}{4\pi}G(| q^2|>>\Lambda_{QCD}) \simeq \frac{4\pi/\beta}{ ln (\pm q^2/\Lambda_{QCD}^2)} \, ,
\ee
where $\pm$ stands for timelike or spacelike $q^2$ respectively. In  the infrared  effective running coupling defined as $ g^2G(0)/(4\pi)=2$, thus the function $G$ is large enough to generate QCD typical CSB and and it provides confinement of quarks for any flavor as well. 
   
\subsection{Model II.}

The second model we use represents a slight modification of the one already considered in  \cite{SABA2008}. 
In this case, the gluon propagator  is expressed through the following spectral representation:
\be  \label{rep}
\frac{g^2}{4\pi}\frac{G(q^2)}{q^2}=\int_0^{\infty} d\nu\, \frac{\rho_g(\nu,\Lambda_{QCD})}{q^2-\nu} \, , 
\ee 
 The function  $\rho_g $ we consider here is a regular smooth function, ensuring thus there is no  pole in the gluon propagator. The main difference comparing to \cite{SABA2008} is that  we drop out the usual Feynman $\ep$ prescription so the principal value integration is understand for the timelike momenta in (\ref{rep}). As far as we are not considering the feedback of complex quark loop the gluon propagator is clearly real for all Minkowski $q^2$.  

In principle the weight function $\rho_g$ can be obtain by solving the gluon gap equation as suggested in \cite{CORNWALL} and further considered recently  in \cite{BIPA2002,AGBIPA2008,Cornwall2008} (up to the changes followed from the   $\ep$ absence).

Following the arguments presented in \cite{Cornwall2008}, it seems that gauge invariant gluon propagator is not "strong" enough to trigger CSB. Very likely, the next leading vertex correction must enforce the kernel to get the expected picture of CSB and confinement. To that point we simplify and make a phenomenological choice of the function $\rho_g$ which provide the desired solution.

   The function we actually use in our numerical study reads 
\be \label{gluonsr}
\rho_g(x)=\frac{\alpha(x)}{\alpha(0)}\frac{\rho_{\alpha}(x)}{x+0.1\Lambda^2}
\ee
%e numerical in the chiral limit.  To achive  a better convergence in the model II a lar
where the function $\alpha(x)$ is calculated through
\bea
\label{repa}
\rho_{\alpha}(x)&=&\frac{4\pi/\beta}{\pi^2-\ln^2{(x/\Lambda^2_{QCD})}} \, ,
\nn \\
\alpha(x)&=&P. \, \int_0^{\infty} d\nu\, \frac{\rho_{\alpha}(\nu)}{x-\nu} \, ,
\eea
where symbol $P.$ stands for Cauchy principal value integration. For the reader familiar with our previous paper on Temporal Euclidean QCD, now gluon form factor is twice weaker there is  factor $2$  omitted in (\ref{repa}), further the infrared cutoff $0.1\Lambda^2$ is introduced in
(\ref{gluonsr}) in order to have finite and smoothed gluon propagator $G/q^2$ in the infrared. Evaluating the integral one can see that  for a very large momenta the gluon propagator is softened by the power of log, i.e.
\be 
\frac{g^2}{4\pi}G(|q^2|>>\Lambda_{QCD}^2) \simeq \frac{4\pi/\beta}{2 ln^2 (\pm q^2/\Lambda_{QCD}^2)}
\ee
(this fact has been overlooked in \cite{SABA2008}). This affects the UV tail of quark mass function, but is quite unimportant for the low the energy  behaviour of the quark mass function $M(0)$.

\section{Solution of the gap equation}
\subsection{QCD light quarks}

The quark gap equation have been solved numerically by the standard method of  iterations. The Gaussian numerical integrator with 600 mesh points was used for  all the numerical integrations.
The angular integration has been performed analytically in the model II, thus we only numerically integrate over the auxiliary variable $\nu$. 
To avoid numerical noise, large density of the mesh points for $q^2\simeq \Lambda, \nu\simeq \Lambda$ was used in the Model II. Furthermore, in order to achieve numerically stability during the run of iteration process,
we gradually decreased current quark mass to  reach the massless limit.
The model I is completely stable, whilst there is a small numerical noise for a low $q^2$
 in the model II.  

Actually, what we have solved was the system of  four real equations for the
functions $R_A,\Gamma_A, R_B, \Gamma_B$. These have been obtained by standard
trace projections (of Dirac matrices), what leads to the coupled system of two equations for the functions $A,B$ in the first step.. After performing the 3d-space Wick rotation, we get the expressions in the Temporal Euclidean space. Further using Eqs. (\ref{sv}) we arrive to the equations for the imaginary and real parts -- Cartesian complex  coordinates of $A,B$ -- the functions $R_A,\Gamma_A, R_B, \Gamma_B$ . Procedure is very  straightforward and we do not list the result here, however the reader can find the equation for $B$ 
in Section \ref{future}, where the  properties of the solution are discussed.

As in the case of perturbation theory, the equations contain UV divergences which require renormalization. The  chiral limit is considered for  both the models. The only renormalization function $A$ is renormalized, the function $B$ is finite in this case.   
In the case of light quarks we assume that the ultraviolet timelike behaviour is similar to the spacelike one, hence we set a  small real renormalized quark mass
 $M(\mu^2)\simeq \Lambda_{QCD}/100$ for a very large timelike $\mu^2$.
In this paper all dimensionfull quantities are scaled by  $\Lambda_{QCD}$,  having
a typical value for $\Lambda_{QCD}=250-350GeV$ we set a few MeV light quarks at $200-300 GeV$ (a given mesh point is chosen for convenience, thus the corresponding value of the renormalization point is  $\mu=920.3\Lambda_{QCD}$ for presented solution here). 
 In this case we subtract the real parts of functions $A$, $B$   setting the renormalized values $ Re A=1,  Re B=\Lambda_{QCD}/100$ at the renormalization scale $\mu$. As in  \cite{SABA2008}, the imaginary part is expected to be finite and therefore not subtracted. This procedure clearly maintains the hermicity of the classical Lagrangian.

Having the  numerical solutions we
calculated  the magnitude and the mass function phase defined as
$M=B/A, M=|M|e^{i\phi_M}$. The appropriate results are presented  in  Fig.1-3. Renormalization wave function $A$ for our models is shown
in  Fig.4. It is approximately real everywhere, the imaginary part portion obtained $\simeq 10^{-8}$ is smaller then estimated numerical error. There is no remarkable difference between the solution of $A$ for $u,d$ quarks and for the exact chiral limit. In the case of model II these two lines are not distinguishable. The absolute values of the mass functions are shown  in Fig.1. In the case of the model II, the  function $|M|$ shows up the maximum at $\simeq 2\Lambda_{QCD}$ where it also cuts the linear function of $p$. This details are better seen in the Fig.2 where we show the infrared details with linear axis scaling.

 As a bonus of our CSB solution we get the pion Bethe-Salpeter
wave function $\chi(P=0,p)\simeq B(p)$ \cite{ROBERTS}, which is the Goldstone boson manifestation of broken chiral symmetry here.

\begin{figure}
\centerline{\epsfig{figure=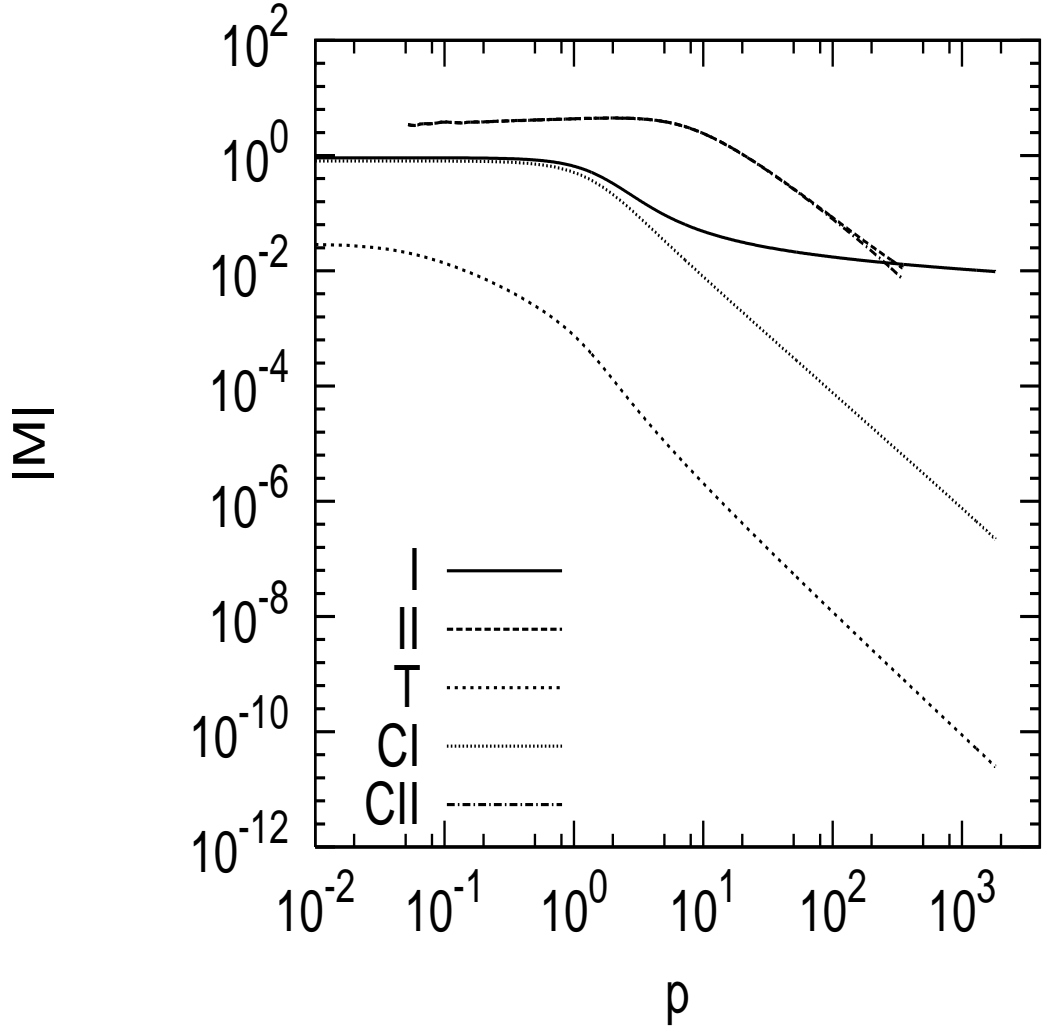,width=15truecm,height=14truecm,angle=0}}
\caption[caption]{Magnitude $|M|$ of the running quark mass function $M=|M|e^{i\phi}$ for modeled QCD I,II, its chiral limit CI,CII. The "Walking Technicolor" T solution is added for for the comparison, scale is $\Lambda_{QCD}=1$ (and $\Lambda_{Tech.}=1$ is set up in this case as well)  } \label{figm}
\end{figure}

\begin{figure}
\centerline{\epsfig{figure=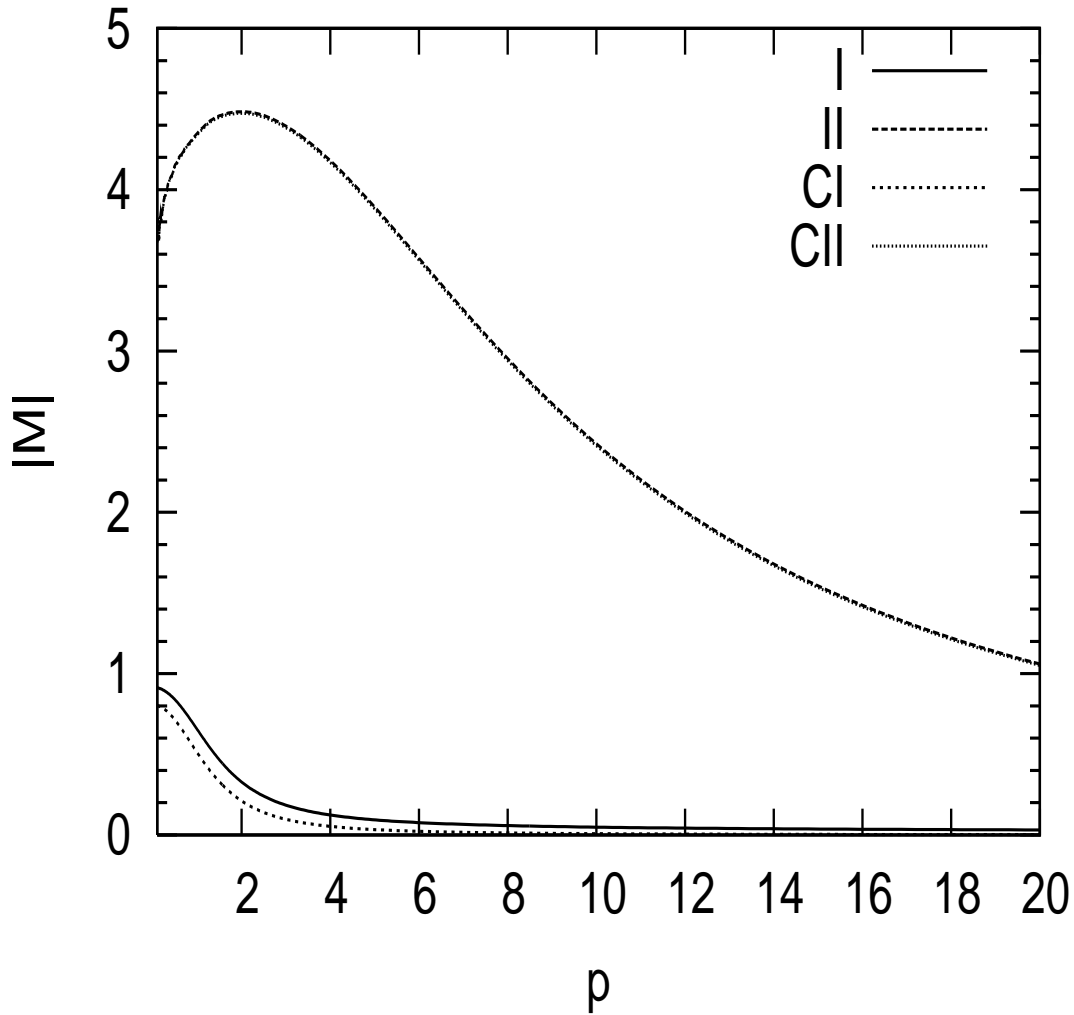,width=15truecm,height=14truecm,angle=0}}
\caption[caption]{I infrared behaviour of the functions $M$ as they are in the Fig. 1, but in with linear axis. The solution for  large $N_f$ is omitted here. } \label{figlm}
\end{figure}

\begin{figure}
\centerline{\epsfig{figure=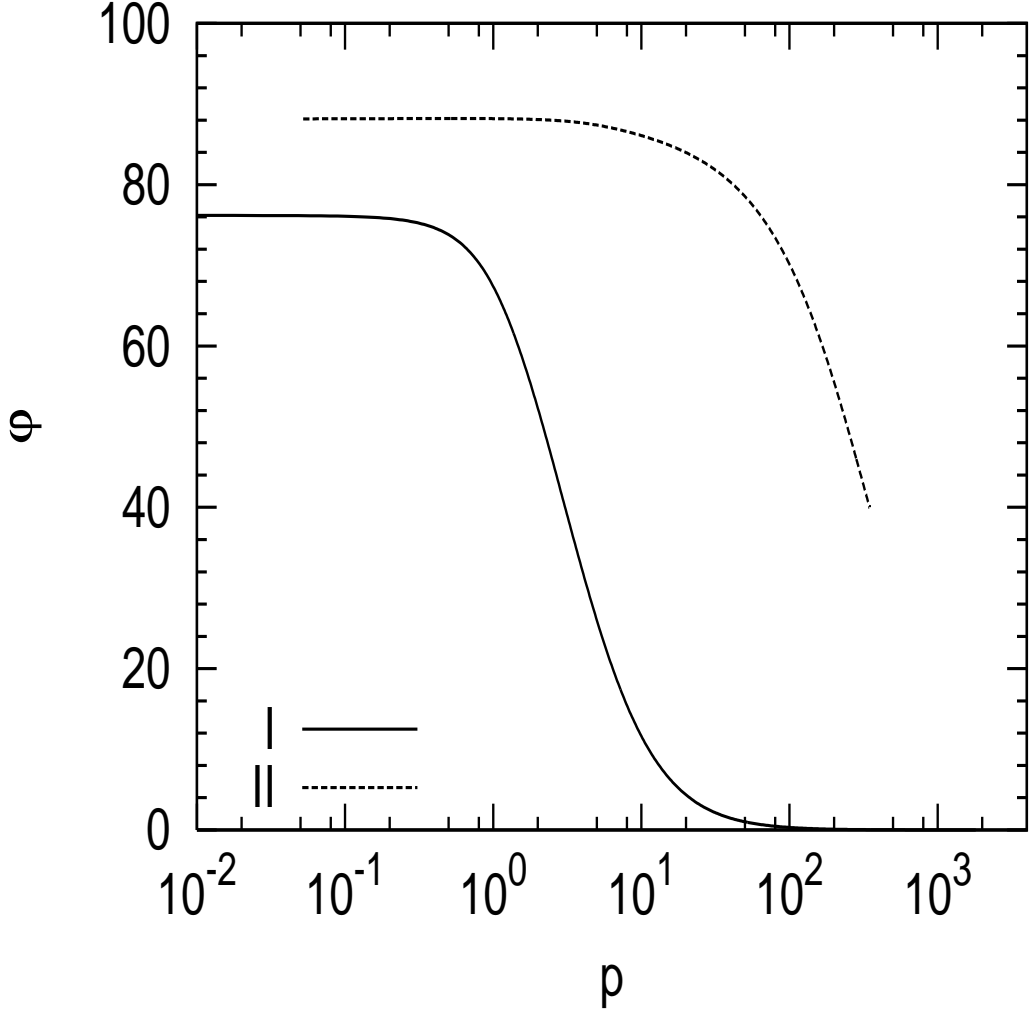,width=15truecm,height=14truecm,angle=0}}
\caption[caption]{Phase $\phi$ of the running quark mass function $M=|M|e^{i\phi}$ for the models I and II respectively, axis momentum is in the units of  $\Lambda_{QCD}$. } \label{figfm}
\end{figure}

\begin{figure}
\centerline{\epsfig{figure=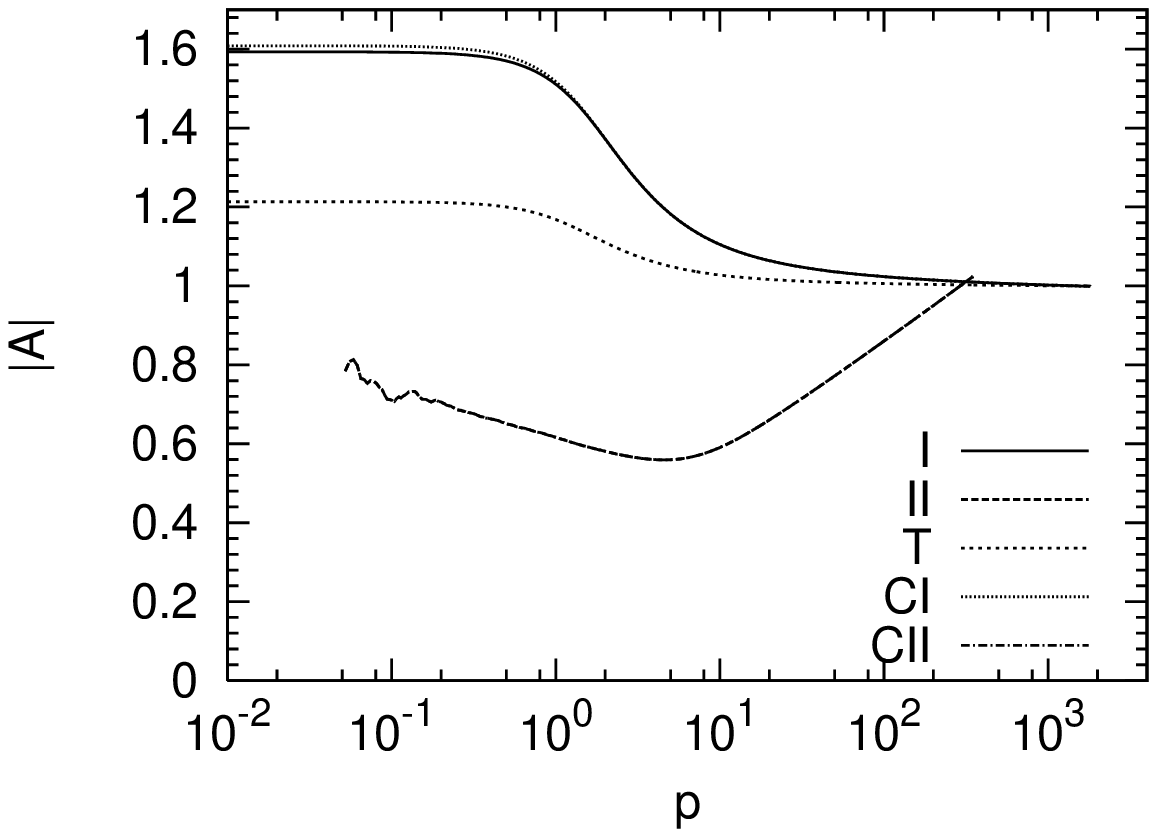,width=15truecm,height=14truecm,angle=0}}
\caption[caption]{Renormalization wave functions for the model I and II. The same is shown for the chiral limit (which are indistinguishable in one case). The "Technicolor" T solution is added for for the comparison, the momentum axis is scaled in  $\Lambda_{QCD}$ and $\Lambda_T$ respectively. } \label{figa}
\end{figure}

Not similarly to previously studied confining theory QED2+1 \cite{SABA2009}, where a quite tiny imaginary part preserves regularity of the time axis propagator, here a huge dynamical generation of imaginary part mass function $M$ is observed. For  pure CSB $(m=0)$ case the observed generated mass is purely imaginary and  the same is approximately valid for the infrared mass when small explicit breaking is considered.  In both cases, the imaginary part vanish at high $p$. The phase $\phi$ of the running quark mass function for the models I and II respectively is shown in Fig \ref{figm}. The chiral limits are always $\phi=\pi/2$ for both models and hence not displayed. Increasing a current quark mass, the phase is decreasing (rather say its absolute value, there are possibly more than one solution for a quark light enough \cite{CLBRW2007}), however from a certain value of the quark current mass it does not disappear at its stay constant from some  $p$. A more complete study on heavy quark confinement will be published elsewhere. Further discussion of the solution, after the following Technicolor digression, will follow in the section \ref{future}.

\subsection{Criticality in large $N_f$ QCD, walking Technicolor and views in ET space}

 QCD is an example of non-Abelian gauge theory with small -two or three- approximately massless fermions. If the number of massless fermions is larger, but the asymptotic freedom is still preserved at some high scale, the theory is conformal in the infrared, which  have been already suggested by the analysis based on two loop  beta function
\cite{CAS1974,BAZA1982} and analyzed for variety of gauge theories (see  \cite{SANKY2009} and references therein).
Non-Abelian quantum filed theory with possibly large but yet supercritical number   of (classically) massless fermion appears as an alternative candidate for electroweak symmetry breaking \cite{HOL1981,AKW1986,AKYA1986}. Typically in  these "walking Technicolor" models, the number of fermions set up the scale $\Lambda_T$ which characterize running of the effective coupling,  which is much larger then fermion mass generation $M(0)<\Lambda_T$. In other words, the conformal infrared nontrivial coupling is adjusted a few percentage above its critical level , say  defined at zero momenta $\alpha_c(0)=g^2/(4\pi)$ , where $g$ is a gauge coupling.
In these theories  the coupling is  "slowly walking" with momentum  in the infrared $p<\Lambda$ where it ensures CSB. The CSB and Technihadrons spectrum in these models have been studied in 
\cite{APRATE1998,APSE1997,GIJA2006,MARO2006}.

Identification of the critical number of the flavors for given theory  requires nonperturbative knowledge  of  infrared behaviour of the running gauge coupling. For $SU(3)$ gauge group 
a recent estimate based on the lattice simulations  gives $8<N_f^c<12$ \cite{APFLNE200,DEPALO2008}.
To model the quark gap equation for large $N_f$ we adjust the infrared gluon form factor 
to be close to  minimal strength necessary to trigger CSB. It is achieved by taking the constant $\beta=1/3$ of the model $I$, which model is particularly suited for this purpose.
We have found the critical couplings are the same in the Standard Euclidean formulation and the one obtained in the Temporal Euclidean space. The obtained CSB solution in  ET space solution is confining one, the mass function $B$ is purely imaginary in his case. The renormalization wave function $A$ remains  real receiving expected smaller corrections than in the case of QCD.

Decreasing infrared coupling furthermore we obtain only trivial solution for the function $B$.

\subsection{Further observation}
\label{future}

Let as consider the gap equation in more details.
As the consequence of our  special model $I$ we get with astonishing accuracy the following relation 

\bea
A(p^2>0)&=&A^{ET}(p^2)=A^{E}(p^2)=A(p^2<0)
\nn \\
B(p^2>0)&=&B^{ET}(p^2)=i B^{E}(p^2)=i B(p^2<0).
\label{symetry}
\eea     

between spacelike and timelike solution. The first is obtained in standard Euclidean formulation, while the second in ET space for pure CSB, $m=0$. A small current mass leads 
to a small deviation only.   

The exact symmetry (\ref{symetry}) is directly visible by the inspection of  the SDE for $B$ which in ET space reads
\bea
B(x)&=&\frac{4}{\pi}\int_{0}^{\infty}dy \frac{B(y)}{A^2(y)y-B^2(y)}\int_{-1}^{1} d z V(x,y,z)
\nn \\
V(x,y,z)&=&\frac{ y \sqrt{1-z^2}}{q^2 \log(e+q^4/\Lambda^4)} \, ,
\eea    
where here $q^2=x+y-2 \sqrt{x y} z$.

For the standard (spacelike) Euclidean formulation the gap equation reads
\be
B(x)=\frac{4}{\pi}\int_{0}^{\infty} dy \frac{B(y)}{A^2(y)y+B^2(y)}\int_{-1}^{1} d z V(x,y,z)
\, .
\ee

The symmetry $\ref{symetry}$ is manifest.
In other words: for the  Minkowski kernel which is even with respect of  argument of the gluon propagator
$G(q^2)=G(-q^2)$, the propagator functions $S_S$ obtained in $E$ and $ET$ spaces differ by the phase factor 
\be
S_s(-p^2)=e^{\pm i\pi/2}S_s(p^2) \, ,
\ee
(we use Minkowski space convention in this expression)
 while the the functions $S_V$`s are identical in both spaces.

The gap equations above  are displayed for small $\beta=1/2$ and  we confirm 
the solution numerically for any $beta$, i.e. the symmetry is kept for  for
increasing $\beta$, when we are gradually leaving the critical point and reaching the value typical for QCD. No other solution was observed numerically for purely dynamical CSB.
We observed that the  non zero  real imaginary parts of $B$ and $A$ are generated  because of nonzero current quark masses, therefore the observed identity (\ref{symetry} are only approximate there in the infrared. In reality,  the full gluon propagator and the quark-gluon vertex are not suppose to be  a simple real and even functions of momenta, but we expect substantial changes due to this. 
We argue here, solutions of quark gap equations for small current masses already presented in the vast amount of the literature are not only Euclidean solutions usually identified with spacelike Minkowski solution, but up to the phase,  they  represent a rough but reliable approximation of the timelike Minkowski solution as well.  

Although we assume  here,  the observed structure is crucial for selfconsistent determination of the Greens function in ET space, the resulting phases  become irrelevant whenever  observables are composed, since the scalar product made from  a given amplitude and its conjugated can only  represents a measurable quantity.

\section{Summary and conclusions}

The first analysis of the light quark gap equations in the Temporal Euclidean space was presented in the paper.
We assume that ET solutions represent  Minkowski solutions at timelike axis as good as the standard Euclidean solutions the Minkowski spacelike one. Behaviour of the solutions were discussed, the main feature of the all solutions is  that the quark mass function is 
predominantly imaginary, providing  confining solution for the quark propagator; such quark 
propagator has no pole nor branch point at real $p^2$ axis.

In  perturbative QCD and in weak coupling theories generally, the well known calculational trick- the Wick rotation-  is under good control. It is a text book knowledge  that timelike solution could be obtainable by an analytical continuation of the result conveniently defined and calculated in spacetime Euclidean space. 
Here, we have assumed no poles and no cuts at a real axis and got complex solution justifying our assumptions. The observed CSB solution in ET space cannot be obtained by analytical continuation of the Euclidean one, it basically  differs by the phase. The meaning of that requires  more deep understanding.

Recently we do not know how to judge and evaluate the quality of an assumptions we made when switching between Minkowski and Euclideans worlds. The all complex space of four momenta is 
not under easy control. On the other side, the  Euclidean metric simplifies things, how well it   approximates   our real Minkowski world can be judge a posterior: The hadrons properties calculated from the QCD Green`S functions are recent (future)  prospectors of quality of Euclidean (Temporal Euclidean) calculations.   
 
 Some possible doubt can follow from unknown prescription of the kernel of considered SDE here. The gluon SDE and related vertices require future selfconsistent analyzes in ET space. It will be certainly  more  evolving but similar task as the one performed here for the quark gap equation alone. On the other side, we do not think that the   improvement towards the exact knowledge of the SDE kernel  can drastically change  qualitative result. The observed complexification phenomena is very universal as  it has happened to two quite  different models. However, going beyond the rainbow ladder is task for  future Temporal Euclidean space study.

At last but not at least the dynamical CSB has been studied near the critical coupling. 
The observed numerical solutions obtained  do not suggest separation of the dynamical CSB from confinement. These phenomena go hand by hand in hypothetical Walking Technicolors models.

%%%%%%%%%%%%%%%%%%%%%%%%%%%%%%%%%%%%%%%%%%%%%%%%%%%%%%%%%%%%%%%%%%%%%%%

\end{document}